\documentclass[11pt]{article}
\usepackage{hyperref}
\pdfoutput=1
\begin{document}
\title{Electro-worming: The Behaviors of \textsl{Caenorhabditis (C.) elegans} in DC and AC Electric Fields}
\author{Han-Sheng Chuang*, David Raizen**, Nooreen Dabbish**, and Haim Bau* \\
\\\vspace{6pt} *Mechanical Engineering and Applied Mechanics,\\\vspace{6pt} **Department of Neurology, \\ University of Pennsylvania, Philadelphia, PA 19104, USA}
\maketitle
%% The abstract (in this file, and that submitted as text to arXiv) should include the exact phrase
%% "fluid dynamics video" or "fluid dynamics videos"
\begin{abstract}
The video showcases how C. elegans worms respond to DC and AC electrical stimulations.  Gabel et al (2007) demonstrated that in the presence of DC and low frequency AC fields, worms of stage L2 and larger propel themselves towards the cathode.  Rezai et al (2010) have demonstrated that this phenomenon, dubbed electrotaxis, can be used to control the motion of worms.  In the video, we reproduce Rezai's experimental results.  Furthermore, we show, for the first time, that worms can be trapped with high frequency, nonuniform electric fields.  We studied the effect of the electric field on the nematode as a function of field intensity and frequency and identified a range of electric field intensities and frequencies that trap worms without apparent adverse effect on their viability.  Worms tethered by dielectrophoresis (DEP) avoid blue light, indicating that at least some of the nervous system functions remain unimpaired in the presence of the electric field. DEP is useful to dynamically confine nematodes for observations, sort them according to size, and separate dead worms from live ones.
\end{abstract}
% main text
\section{Introduction}
%% The {\em hyperref} package is used to make links to the videos.
%% The format is: \href{URL of video}{name that will appear in the text}
The fluid dynamics video is
\href{http://ecommons.library.cornell.edu/bitstream/1813/17513/2/Electroworming_LD.mpg}{Video1} \\
\\
We study the effect of electric field magnitude and frequency on C. elegans. Low magnitude, DC electric fields have been used before to guide the motion of the wild-type nematode C. elegans$^{1,2}$. However, the worms appear to be oblivious to uniform electric fields when the field frequency exceeds several tens of hertz. In contrast, nonuniform, moderate intensity, high frequency ($>$100 kHz) AC fields trap worms at the location of the highest field intensity. With certain electrode arrangements, only the worm's tail is immobilized. When the electric field intensity is moderate or high and the frequency is low (about 1-100 kHz), the worm is eventually injured, paralyzed, or electrified depending on the applied electric field magnitude. This is the first demonstration of dielectrophoretic trapping of an animal. The effects of the electric field intensity and frequency on the worm are recorded in a "phase diagram."
\\
The effect of the DC field on the worm is illustrated in a video featuring a conduit made in a PDMS slab.  The worm swims towards the cathode.  When the electric field polarity is reversed, so is the worm's direction of motion.
\\
The effects of nonuniform electric fields on worms are studied with a pair of spiked electrodes patterned on a glass slide and set at various distances apart.  The glass slide caps a trench molded in PDMS to form a 118 $\mu$m tall and 300 $\mu$m wide conduit. The total length of the conduit is 17 mm. The conduit is initially filled with deionized (DI) water with electric conductivity of $1\times10^{-3}$ S/m.  Worms from a synchronous culture are transferred from the culture dish, placed in the inlet of the microfluidic conduit, and propelled gently to the location of the electrodes with the aid of a syringe.  The behavior of individual worms as a function of electric field intensity and frequency is monitored with an optical microscope.  The flow field induced by the worm is monitored by seeding the liquid with fluorescent particles and tracking their positions as functions of time to construct the velocity vector field.  When the worm is smaller than the gap between the electrodes, typically the worm becomes anchored to the electrode by its tail while its more energetic head moves vigorously.
\\
We trapped worms using electric field intensities and frequencies which appeared to leave the worms unharmed. Trapped worms, after release, tend to function as untrapped worms for many hours. The anchored worm appears to produce swimming motion similar to that of an untrapped worm in the absence of an electric field.  Since the trapped worm cannot move, the liquid around it is propelled backwards by the worm's undulatory motion.  Thus, the anchored worm acts as a pump.  The worm's motion induces vortices in the flow and is likely to provide effective stirring; thus the anchored worm can also act as a stirrer.  
\\
Like untrapped worms, the anchored worms exhibit photophobicity - avoiding blue light. Thus, blue light can be used to exert control on the worm's motion.  Furthermore, the blue-light avoidance indicates that (at least) the light-sensitive neurons are not adversely impacted by the electric fields. 
\\
Two worms anchored in close proximity synchronized their motion.  Although the synchronization mechanism is not fully understood, it is most likely that hydrodynamic stimuli played a role in transmitting information from one worm to the other. 
\\
Dielectrophoresis can be used, among other things, to sort worms by size, to temporarily anchor worms to enable their characterization and study, and to use worms to induce fluid motion (worm-pump) and mixing (worm-stirrer).
\\
\section{References}
\begin{enumerate}
\item
P. Rezai, A. Siddiqui, P. R. Selvaganapathy and B. P. Gupta, Appl. Phys. Lett., 2010, 96, 153702.
\item
C. V. Gabel, H. Gabel, D. Pavlichin, A. Kao, Clark D.A. and A. D. T. Samuel, J. Neurosci., 2007, 27, 7586-7596.
\end{enumerate}
\end{document}